\documentclass[]{emulateapj}

\usepackage{graphics,graphicx,xspace,natbib,amssymb,color}
\usepackage{amsmath}

\shorttitle{Color Gradients \& Half-mass Radii from $0\lesssim\lowercase{z}\le 2.5$}
\shortauthors{Suess et al.}

\begin{document}

\title{Half-mass radii of quiescent and star-forming galaxies evolve slowly from $0\lesssim\lowercase{z}\le 2.5$: implications for galaxy assembly histories\altaffilmark{*}}
\author{Katherine A. Suess\altaffilmark{1}, Mariska Kriek\altaffilmark{1}, Sedona H. Price\altaffilmark{2}, Guillermo Barro\altaffilmark{3}} 

\altaffiltext{*}{This work is based on observations taken by the CANDELS Multi-Cycle Treasury Program and the 3D-HST Treasury Program with the NASA/ESA {\it HST}, which is operated by the Association of Universities for Research in Astronomy, Inc., under NASA contract NAS5-26555.}
\altaffiltext{1}{Astronomy Department, University of California, Berkeley, CA 94720, USA}
\altaffiltext{2}{9Max-Planck-Institut f{\"u}r extraterrestrische Physik, Postfach 1312, Garching, 85741, Germany}
\altaffiltext{3}{Department of Physics, University of the Pacific, 3601 Pacific Ave, Stockton, CA 95211, USA}
\email{suess@berkeley.edu}

\begin{abstract}
We use high-resolution, multi-band imaging of $\sim$16,500 galaxies in the CANDELS fields at $0\lesssim z\le 2.5$ to study the evolution of color gradients and half-mass radii over cosmic time. We find that galaxy color gradients at fixed mass evolve rapidly between $z\sim2.5$ and $z\sim1$, but remain roughly constant below $z\sim1$. This result implies that the sizes of both star-forming and quiescent galaxies increase much more slowly than previous studies found using half-light radii. The half-mass radius evolution of quiescent galaxies is fully consistent with a model which uses observed minor merger rates to predict the increase in sizes due to the accretion of small galaxies. Progenitor bias may still contribute to the growth of quiescent galaxies, particularly if we assume a slower timescale for the minor merger growth model. The slower half-mass radius evolution of star-forming galaxies is in tension with cosmological simulations and semi-analytic galaxy models. Further detailed, consistent comparisons with simulations are required to place these results in context. 
\end{abstract}


%------------------------------------------------------------------------------------------------------------------------------------------------------------------------
\section{Introduction}

Examining how star-forming and quiescent galaxies grow in size over cosmic time can provide clues about galaxy evolution. In combination with other measured properties, sizes inform studies of the mass assembly history of galaxies \citep[e.g.,][]{bezanson09,naab09}, the connections between galaxies and their host dark matter halos \citep[e.g.,][]{mo98,kravtsov13,somerville17,jiang19}, how galaxy populations change \citep[e.g.,][]{carollo13,poggianti13} and disks build up over time \citep[e.g.,][]{vandokkum13}. 

Star-forming galaxies seem to grow more slowly than zeroth-order expectations for dark matter halo growth \citep[][]{mo98}, perhaps due to evolving halo spin parameters \citep[e.g.,][]{somerville08} or feedback processes \citep[e.g.,][]{dutton09}. Recent abundance matching studies including this additional physics have found that the observed sizes of disky star-forming galaxies are indeed proportional to the virial radii of their host halos \citep[e.g.,][]{kravtsov13,huang17,somerville17}. Cosmological simulations have also found good agreement between observed and modeled star-forming galaxy size evolution \citep{furlong17,genel18}. 

Quiescent galaxies appear to grow in size incredibly rapidly, especially at high redshift \citep[e.g.,][]{daddi05,vandokkum08,damjanov09,szomoru10,damjanov11,vanderwel14}. Two theories have arisen to explain this rapid size evolution: in the `inside-out growth' scenario, individual quiescent galaxies grow via minor mergers \citep[e.g.][]{bezanson09,naab09,hopkins09,vandesande13}; in the `progenitor bias' scenario, the median size of the quiescent population increases with time because galaxies that quench at later times are larger \citep[e.g.][]{vandokkum01,carollo13,poggianti13}. The apparent size growth of quiescent galaxies could be explained by individual growth, population growth, or a combination of both processes.

All of these observational studies measure galaxy sizes from stellar light profiles. However, radial $M/L$ variations cause a galaxy's light profile to differ from its mass profile. Therefore, half-light radii are {\it biased tracers} of galaxy mass distributions if $M/L$ gradients are present. In \citet{suess19}, we measured the half-mass radii of $\sim7000$ galaxies from multi-band high-resolution imaging and showed that the strength of color gradients in both star-forming and quiescent galaxies evolves between $z=1.0$ and $z=2.5$. As a result, the evolution of galaxy half-mass radii differs from the previously-reported half-light radius evolution. In this Letter, we extend our previous analysis to lower redshifts. Using this uniform sample of half-mass radii from $0\lesssim z\le 2.5$, we analyze and discuss how star-forming and quiescent galaxies grow in size over cosmic time.

Throughout this Letter, we assume a cosmology of $\Omega_{\rm{m}}=0.3$, $\Omega_\Lambda=0.7$ and $h= 0.7$; all radii are non-circularized measurements of the major axis.

%------------------------------------------------------------------------------------------------------------------------------------------------------------------------
\section{Sample, Methods, \& Galaxy Half-Mass Radii}

\begin{figure*}
    \centering
    \includegraphics[width=.7\textwidth]{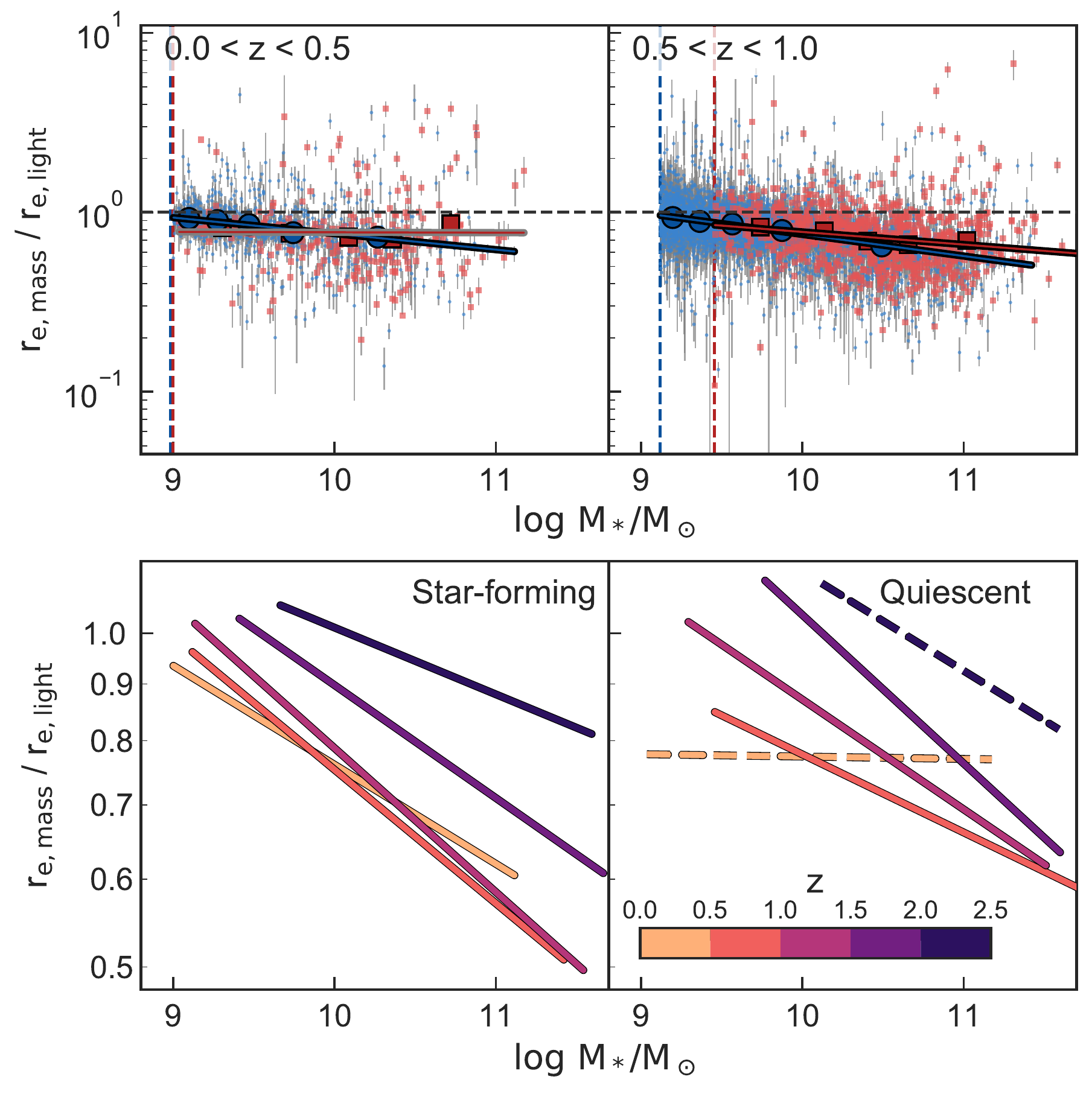}
    \caption{Top: color gradient strength as a function of stellar mass for two redshift intervals. Small light blue points and light red squares show individual star-forming and quiescent galaxies; large blue points and red squares show a running median. The blue and red lines show best-fit linear relations to each trend; lines are outlined in black if the slope of the relation is inconsistent with zero, and outlined in grey if the slope is consistent with zero. Dashed vertical blue and red lines show the mass completeness of the star-forming and quiescent samples. Bottom: best-fit relation between color gradient strength and stellar mass for star-forming (left) and quiescent (right) galaxies as a function of redshift. Dashed lines represent fits whose slopes are consistent with zero. We see clear redshift evolution in the best-fit color gradient-mass relations for both quiescent and star-forming galaxies.}
    \label{fig:sample}
\end{figure*}

In this Letter, we present the evolution of galaxy half-mass radii from $z=2.5$ to $z\sim0$. 
From $1.0\le z\le2.5$, we use the sample of galaxy half-mass radii presented in \citet{suess19}. This high-redshift sample consists of 7,006 galaxies selected from the ZFOURGE photometric catalog \citep{straatman16} to have $\log{M_*/M_{\odot}}>9.0$, S/N$_{\rm{K}}\ge 10$, a use flag equal to one, a match in the 3D-HST catalog \citep{brammer12,momcheva16,skelton14}, and a convergent \texttt{GALFIT} fit \citep{peng02,vanderwel14}.

Here, we expand our previous work by presenting the half-mass radii of an additional 9,543 galaxies at $z\le1.0$. This low-redshift sample consists of all galaxies in the 3D-HST photometric catalog with $z\le1.0$, $\log{M_* /M_{\odot}}>9.0$, S/N$_{\rm{F160W}}\ge 10$, a use flag equal to one, and a convergent \texttt{GALFIT} fit. These selection criteria are equivalent to those of the higher-redshift sample presented in \citet{suess19}, but using the 3D-HST catalogs as opposed to the ZFOURGE catalogs. This allows us to include galaxies from the AEGIS and GOODS-N fields (not included in ZFOURGE) and better sample the $z\lesssim0.5$ universe. Furthermore, we note that recovered galaxy properties at $z<1$ do not change significantly with the inclusion of the ZFOURGE data: at these low redshifts, the medium-band filters in ZFOURGE no longer sample the Balmer break. 

We calculate the half-mass radii of galaxies in the low-redshift sample following the methods of \citet{suess19}. In brief: we calculate aperture photometry in elliptical annuli for each galaxy in each band of PSF-convolved {\it HST} imaging \citep{skelton14}, then use FAST \citep{kriek09} to model the resulting spatially-resolved spectral energy distributions (SEDs). This method produces an as-observed $M/L$ profile for each galaxy. We correct for the effects of the point spread function with a simple forward modelling technique which assumes that that the intrinsic $M/L$ profile is a power-law function of radius. We then use the best-fit intrinsic $M/L$ profile in conjunction with the galaxy's light profile and the point spread function to find the half-mass radius. Full details of this method and comparisons with other techniques for measuring half-mass radii are presented in \citet{suess19}. \citet{suess19} also shows that the half-mass radii recovered using our technique are not significantly biased by the galaxy's half-light radius, stellar mass, or redshift.

In Figure~\ref{fig:sample}, we show the strength of galaxy color gradients (i.e., $r_{\rm{e,mass}}/r_{\rm{e,light}}$) as a function of stellar mass for two redshift slices in our new low-redshift sample. We  divide the sample into star-forming and quiescent populations using the UVJ diagram and the quiescent definition from \citet{whitaker12_psb}. In \citet{suess19}, we found that there was a significant trend between color gradient strength and stellar mass for galaxies at $1.0\le z\le 2.5$. Here, we find that this trend continues to lower redshifts. The only exception is in the lowest-redshift quiescent bin, where our sample is quite small. In the lower panels of Figure~\ref{fig:sample}, we also show how the best-fit relation between $r_{\rm{e,mass}}/r_{\rm{e,light}}$ and stellar mass varies as a function of redshift for our full $0\lesssim z\le 2.5$ sample.

\begin{figure}
    \centering
    \includegraphics[width=.48\textwidth]{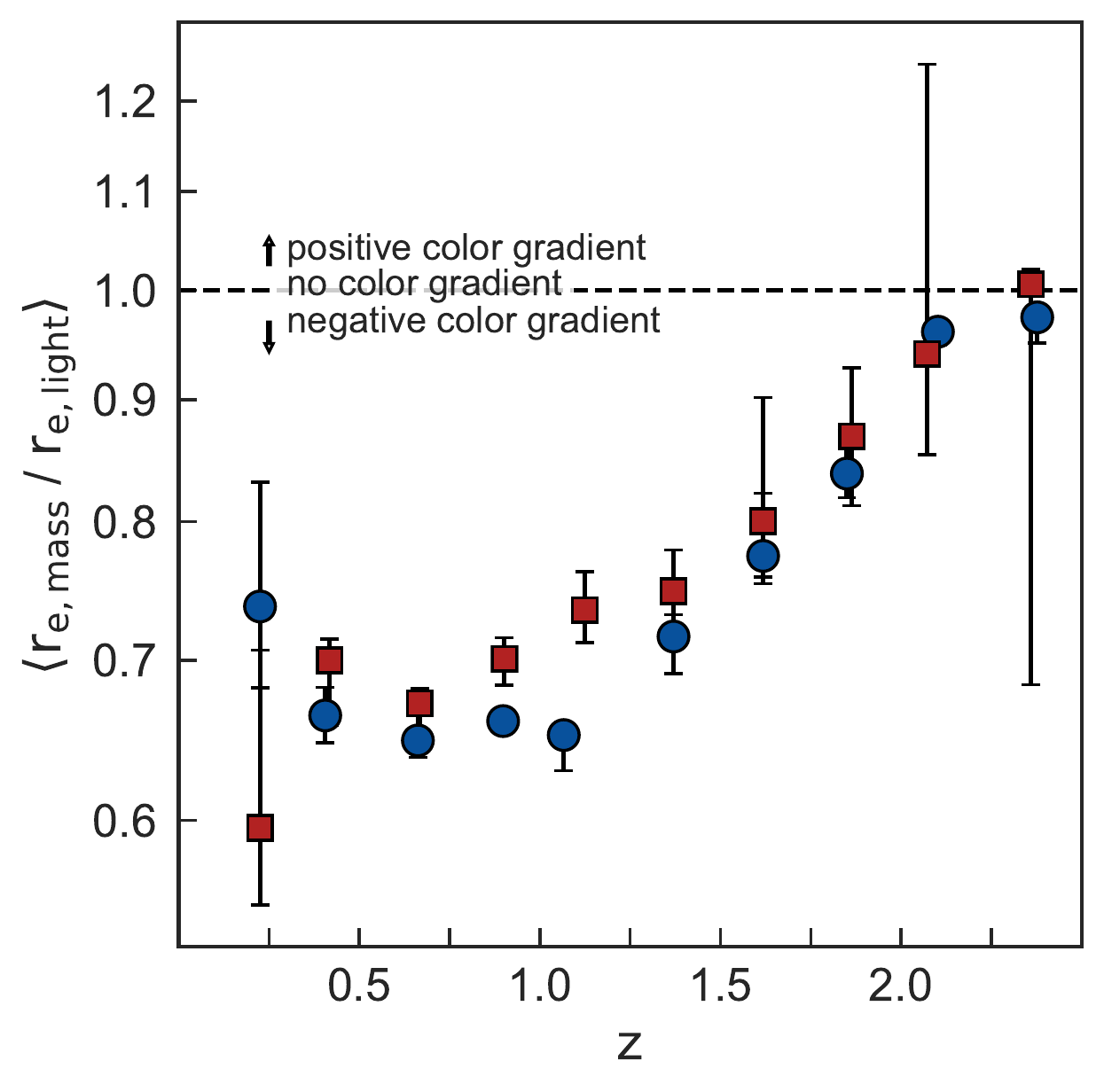}
    \caption{Median half-mass to half-light radius ratio for star-forming and quiescent galaxies (blue circles and red squares) as a function of redshift. $r_{\rm{mass}}/r_{\rm{light}}$ traces the strength of radial color gradients; values less than one indicate negative color gradients, where the center of the galaxy is redder than the outskirts. Only galaxies with $\log{M_*/M_\odot}>10.1$, where our sample is complete, are included. Error bars show the central 68\% of 500 bootstrap samples. The strong color gradient evolution previously observed at $z\gtrsim1$ appears to flatten at $z\lesssim1$.}
    \label{fig:rRatio}
\end{figure}

In Figure \ref{fig:rRatio}, we show the median strength of color gradients for all galaxies with $\log{M_*/M_\odot}>10.1$, the completeness limit of our full sample. In \citet{suess19}, we showed that the strength of galaxy color gradients decreases sharply between $z\sim2.5$ and $z\sim1$. Here we find that this evolution may flatten below $z\sim1$; $r_{\rm{e,mass}}/r_{\rm{e,light}}$ for both quiescent and star-forming galaxies remains roughly constant at a value of $\sim0.7$. 

%------------------------------------------------------------------------------------------------------------------------------------------------------------------------
\section{The growth of quiescent galaxies is consistent with minor merger predictions}

Over the past decade, numerous studies have assessed how merger-driven inside-out growth and progenitor bias can contribute to the size evolution of quiescent galaxies \citep[e.g.,][]{bezanson09,vandesande13,hopkins09,carollo13,poggianti13,williams17,damjanov19}. Several studies--- notably \citet{newman12} and \citet{belli15}--- attempt to tease out the relative contributions of these two growth mechanisms. \citet{newman12} studied the merger rates of galaxies in the CANDELS survey and calculated the resulting effect on galaxy sizes. They found that while minor mergers could explain the relatively slow growth in half-light radii at $z\sim1$, minor mergers were {\it insufficient} to explain the much more rapid growth at $z\sim2$. In a similar vein, \citet{belli15} infer the star formation histories of quiescent galaxies and reconstruct the size evolution of the quiescent population. They find that both individual growth and progenitor bias must contribute (in roughly equal proportions) to the observed increase in the half-light radii of quiescent galaxies. Both individual and population growth seem to be necessary to explain the increase in quiescent galaxy sizes, especially at early times when half-light radii grow very rapidly. However, both of these studies used half-light radii to study quiescent size growth; here, we use half-{\it mass} radii to account for the bias due to evolving color gradients.

\begin{figure}
    \centering
    \includegraphics[width=.48\textwidth]{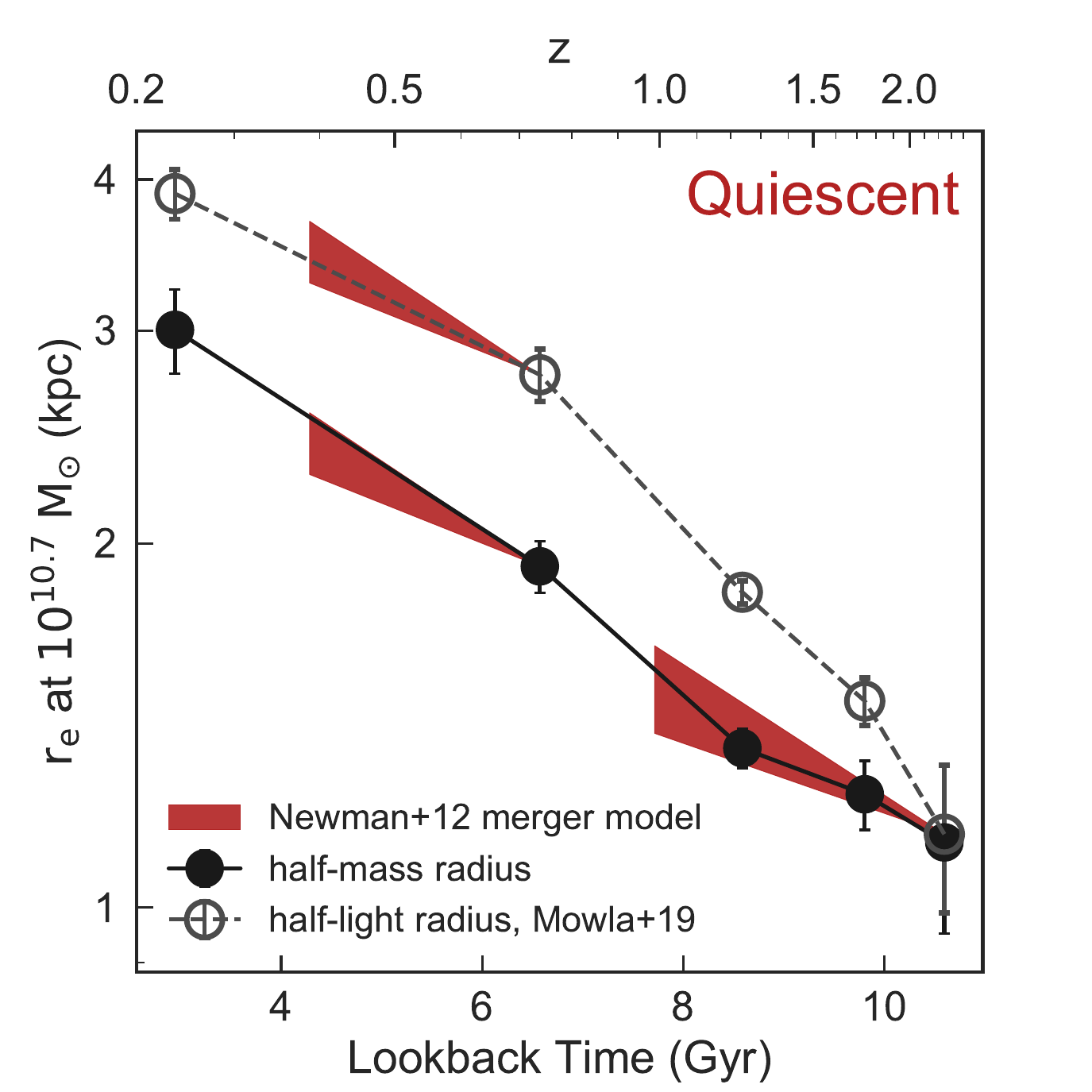}
    \caption{Half-light radius (open grey points) and half-mass radius (filled black points) of quiescent galaxies at $10^{10.7} M_\odot$ as a function of lookback time. Red shaded regions show the expected growth via minor mergers using the \citet{newman12} models, assuming a merger timescale of $\tau_e=1.0$~Gyr. Minor mergers alone are sufficient to explain the growth of half-mass radii.}
    \label{fig:qui_growth}
\end{figure}

In Figure~\ref{fig:qui_growth}, we show the evolution of the half-light and half-mass radii of quiescent galaxies at $M_*=10^{10.7}M_\odot$. The grey curve shows the half-light radius evolution, with data taken from \citet{mowla18}. The black curve shows the evolution of quiescent half-{\it mass} radii at the same stellar mass. The half-mass radius data points are calculated by multiplying our best-fit relation for $r_{\rm{e,mass}}/r_{\rm{e,light}}-M_*$ (Figure~\ref{fig:sample}) by the \citet{mowla18} $r_{\rm{e,light}}-M_*$ relations, as described in more detail in \citet{suess19}. While the half-mass and half-light radii of massive quiescent galaxies are nearly equal at $z=2.25$, half-mass radii grow much less rapidly than half-light radii do.      

We compare this size evolution to the \citet{newman12} predictions for quiescent galaxy growth via minor mergers. Like \citet{newman12}, we assume a merger timescale $\tau_e=1.0$~Gyr. The shaded red region in Figure~\ref{fig:qui_growth} shows this model; the width represents the measured uncertainty in the merger fraction. Between $z=2.25$ and $z=1$, this data-based model predicts that minor mergers cause galaxy sizes to grow by $(0.12\pm0.04)$~dex. 
As identified by previous works, minor mergers alone cannot explain the 0.28~dex increase in galaxy half-light radii found over the same redshift range. However, the growth in half-mass radii (which we measure to be 0.14~dex) is fully consistent with a model that includes minor mergers {\it alone}. No additional mechanism for size growth at high redshift is necessary to explain the observed half-mass radius evolution. 
Between $z=0.75$ and $z=0.4$, the \citet{newman12} model predicts a size increase of $(0.10\pm0.04)$~dex due to minor mergers. Observed half-light radii grow by 0.10~dex, and observed half-mass radii grow by 0.14~dex; both are consistent with the data-based merger growth model. As discussed in \citet{suess19}, our observed color gradient evolution at high redshift also supports an inside-out growth or two-phase formation scenario \citep{naab09,oser10}: minor mergers deposit bluer stars at the outskirts of the galaxy, creating negative color gradients. As redshift decreases and galaxies experience more minor mergers, these color gradients become stronger.

We note that in Figure~\ref{fig:qui_growth} we invoke a fast merger timescale of $\tau_e =1.0$~Gyr. A slower merger timescale would decrease the amount of size growth that minor mergers can account for. In this case, there would still be additional growth in quiescent half-mass radii that must be accounted for by progenitor growth. A careful study of the half-mass radii of the smallest quiescent galaxies over time \citep[e.g.,][]{carollo13}, potentially in combination with stellar abundance studies \citep[e.g.,][]{kriek16,kriek19}, is required to fully understand how much progenitor bias contributes to quiescent half-mass radius growth.

%------------------------------------------------------------------------------------------------------------------------------------------------------------------------
\section{The slow growth of star-forming galaxies is inconsistent with simulations}
\label{sec:sf_growth}

\begin{figure}
    \centering
    \includegraphics[width=.48\textwidth]{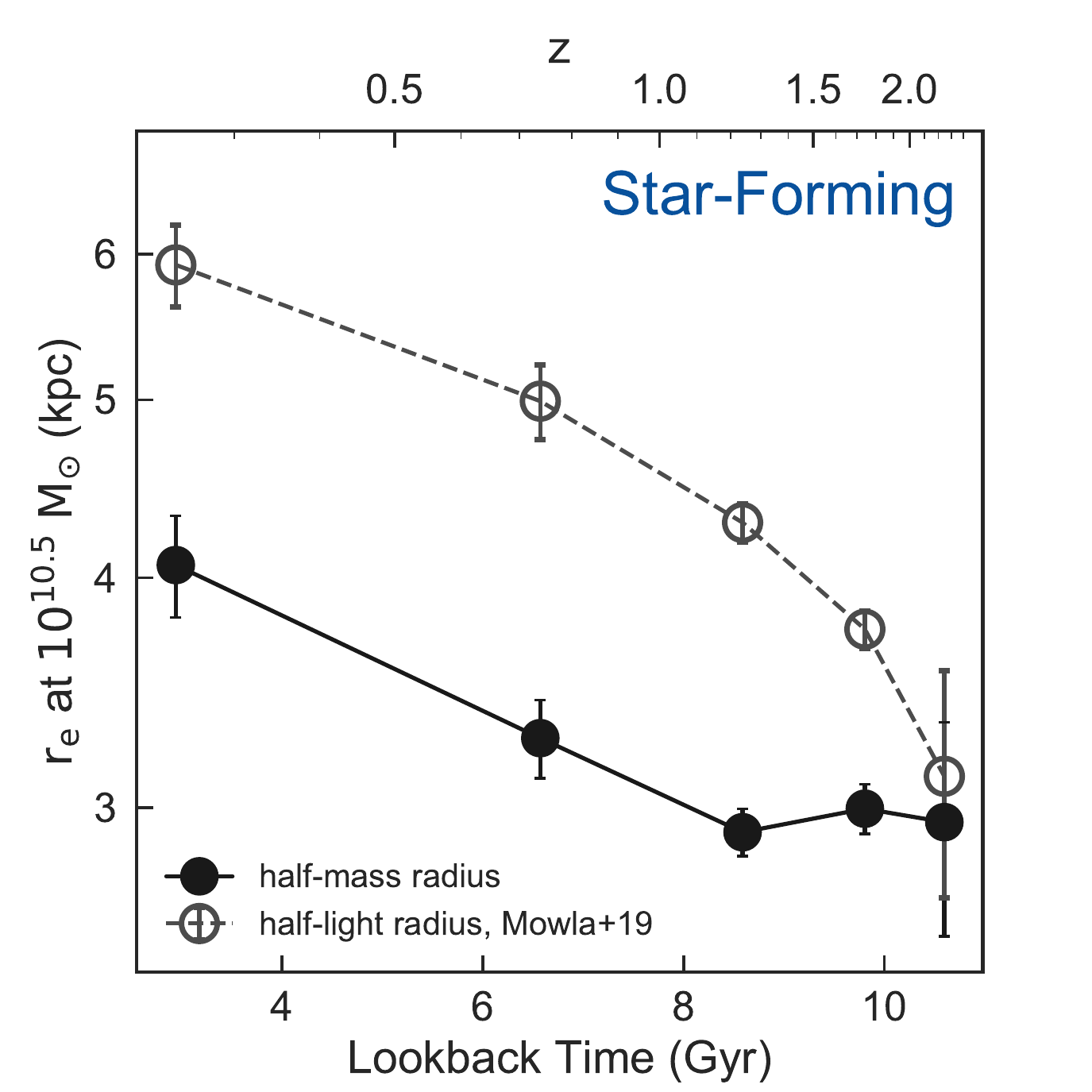}
    \caption{Half-light radius (open grey points) and half-mass radius (filled black points) of star-forming galaxies at $10^{10.5}M_\odot$ as a function of lookback time. Half-mass radii evolve much less rapily than half-light radii.}
    \label{fig:sf_growth}
\end{figure}

We now examine the growth of star-forming galaxies when including the effects of color gradients. Figure \ref{fig:sf_growth} shows how the half-light and half-mass radii of star-forming galaxies evolve. The overall picture is similar to quiescent galaxies: star-forming galaxies have roughly equal half-mass and half-light radii at early times; however, half-mass radii evolve much more slowly than half-light radii. This shallow size evolution is supported by spatially-resolved spectral studies: low- and intermediate-mass star-forming galaxies at $z\sim2$ have nearly flat sSFR profiles, implying self-similar growth and thus slow size evolution \citep{tacchella15b,tacchella15,nelson16}.

Many cosmological simulations and semi-analytic models predict the size evolution of star-forming galaxies based on their mass distributions. Interestingly, the evolution of these simulated half-mass radii generally matches the observed half-{\it light} radius evolution, not the slower half-mass radius evolution we show in Figure~\ref{fig:sf_growth}. Hence, the simulations predict a much stronger evolution in star-forming half-mass radii than our observations. 
For example, \citet{furlong17} show that the predicted half-mass radii of disk galaxies in the EAGLE simulations are consistent with observed half-light radii within $\lesssim0.1$~dex; their modeled half-mass radii show a similar redshift evolution as the \citet{vanderwel14} half-light radius observations. The half-mass radii of galaxies in Illustris-TNG also evolve at a similar rate as observed half-light radii \citep{genel18}. The semi-analytic model of \citet{dutton11} predicts the evolution of disk sizes as well as the strength of color gradients in disk galaxies; their half-mass radius evolution is generally consistent with observations of half-light radii, not with our observed half-mass radii. 

Several physical processes could account for this discrepancy between simulated and observed half-mass radii. \citet{dutton09} found that including feedback in semi-analytic models slows down the expected size evolution of disk galaxies. If the effects of feedback were slightly stronger--- or the physics of feedback was slightly different--- it may be possible to further slow the modeled evolution of star-forming half-mass radii and bring them in line with our observations. Higher merger rates at high redshift could also weaken the evolution of simulated disk sizes. 
Additional explanations for this discrepancy--- which could apply to measurements of half-mass radii for both star-forming and quiescent galaxies--- are discussed in more detail below. 

\section{Discussion}

In general, most modern simulations predict both the mass distributions {\it and} the light distributions of model galaxies, then report half-light radii which can be directly compared to observations. The measured half-light radius evolution of both star-forming and quiescent galaxies from these simulations is in good agreement with observations: the half-light radii typically agree to within $0.1-0.25$~dex \citep{somerville08,dutton11,price17,genel18}. This is particularly interesting in light of our results: first, if the modeled half-light radii agree with observations while the modeled half-mass radii do not (Section~\ref{sec:sf_growth}), then both the half-mass radii {\it and} the color gradients in the simulations may differ from what we observe; second, because these simulations {\it already} include color gradients, these simulated color gradients can now be directly compared with our new color gradient measurements.

To date, very few studies have discussed color gradients in simulated galaxies.
\citet{dutton11} used a semi-analytic model of disks evolving within dark matter halos to compare observed and simulated galaxy scaling relations. They report both half-mass radii and half-light radii in several different filters. Color gradients in their simulated galaxies seem to evolve less rapidly than our observations, decreasing by only $\sim0.1$~dex between $z=2.5$ and $z=0.25$; over the same redshift range, our measured color gradients decrease by $\sim0.3$~dex. The \citet{dutton11} model did not, however, account for the effects of dust or mergers. Both dust and mergers can affect the strength of color gradients in galaxies, and their inclusion in the \citet{dutton11} models would likely change how their simulated color gradients evolve. There are no studies which examine color gradient strength or evolution in modern high-resolution cosmological simulations. Such studies would provide an invaluable comparison to our recent observations, and allow us to more fully contextualize our findings in this paper.

Such a comparison of simulated and observed half-mass radii and color gradients--- while necessary--- will also be difficult.
There are a host of choices to make when analyzing simulations that complicate a direct comparison with observations. Aperture effects can bias galaxy sizes by $\sim0.1$~dex, and varying the viewing angle can change the inferred half-light size by $\sim0.1-0.2$~dex \citep{price17}. Even the total stellar mass of the galaxy--- and thus its location in the mass-size diagram--- can change based on how far out the mass profile is integrated \citep[causing variations of $0.1-0.2$~dex;][]{genel18} and/or differences in the stellar mass loss prescription used by the simulation and by the stellar population synthesis modeling of the observations \citep[e.g.,][]{price17}. Selection effects may also play a role: our observations compare the sizes of star-forming and quiescent galaxies at fixed mass, but if these populations evolve differently in the simulations (e.g., if galaxies quench at a different rate), then the galaxy populations will differ between the observations and simulations. To perform a fair comparison between observations and simulations, it is thus necessary to analyze mock observations of the simulations using the {\it same techniques} that are used for the real observations \citet[see][]{price17}.

Finally, we note that measuring color gradients and half-mass radii from observed data is difficult; we refer the reader to \citet{suess19} for a full discussion of the sources of possible biases in our measurements. In particular, it is difficult to account for the most highly dust-obscured star formation with our methods. Measurements at longer wavelengths \citep[with ALMA, e.g.][]{barro18}, are crucial to understand the mass profiles of such obscured systems.

%------------------------------------------------------------------------------------------------------------------------------------------------------------------------
\section{Summary}
\label{sec:discussion}

In this Letter, we extended the \citet{suess19} analysis of color gradients and half-mass radii of both star-forming and quiescent galaxies to $z\sim0$. This data set, together with our previous results, represents the largest collection of galaxy half-mass radii at $z>0$ in the literature. This large sample of half-mass radii, calculated in a uniform way across a wide range of redshifts and stellar masses, allows us to conduct a detailed examination of how the half-mass sizes of galaxies grow over cosmic time.

We find that color gradients in both star-forming and quiescent galaxies evolve between $z=2.5$ and $z\sim0$. When taking these color gradients into account, the sizes of both star-forming and quiescent galaxies grow much less rapidly than previously found. For quiescent galaxies, the evolution of half-mass radii is fully consistent with the expected growth due to minor mergers {\it alone} (Figure~\ref{fig:qui_growth}) based on the observed merger rates of \citet{newman12} and assuming a relatively fast merger timescale of 1~Gyr. A slower merger timescale would allow additional room for progenitor bias to contribute to the growth of quiescent galaxies. For star-forming galaxies, the evolution of half-mass radii is much slower than predicted by cosmological simulations and semi-analytic models, and raises questions as to the physical mechanisms responsible for this slow size growth. 

Further work is required to reconcile these new observational results with the apparent consensus in the literature. For quiescent galaxies, the magnitude of growth due to progenitor bias is still not fully understood; this may be addressed by detailed examinations of the half-mass radii of the smallest galaxies over cosmic time \citep[as in][]{carollo13,poggianti13}. The slow size evolution of star-forming galaxies is more challenging to address, as it is in tension with available theoretical predictions. Additional work to understand the effects of feedback and mergers on half-mass radii and color gradients is required to address this discrepancy. Finally, studies that directly compare half-mass and half-light radii in cosmological simulations in a consistent manner as the observations \citep[e.g.][]{price17} are essential to understand the origins and impacts of our observed evolving color gradients.

\acknowledgements
We would like to thank Jenny Greene, Pieter van Dokkum, Tom Zick, and the anonymous reviewer. This work is funded by grant AR-12847, provided by NASA though a grant from the Space Telescope Science Institute (STScI) and by NASA grant NNX14AR86G. This material is based upon work supported by the National Science Foundation Graduate Research Fellowship Program under grant No. DGE 1106400.

\bibliographystyle{aasjournal}
\bibliography{lowzbib}

\end{document}